\documentclass[10pt]{amsart}

\usepackage{tikz}
\usepackage{amssymb,amsmath,latexsym,graphicx}
\usepackage{charter,eucal}
\usepackage{amssymb}
\usepackage{amscd, relsize,soul,tikz,color,xcolor}

\setul{}{1.1pt}

\usepackage[linktoc=page,colorlinks,linkcolor={red!80!black},citecolor={red!80!black},urlcolor={blue!80!black}]{hyperref}

\newtheorem{theorem}{Theorem }[section]
\newtheorem{lemma}[theorem]{Lemma}
\newtheorem{remark}[theorem]{Remark}
\newtheorem{corollary}[theorem]{Corollary}

\newtheorem{question}[theorem]{\textsc{Question}}
\newtheorem{conjecture}[theorem]{\textsc{Conjecture}}
\newtheorem{definition}[theorem]{\textsc{Definition}}

\def\1{\mathrel{\mathbf{1}}}

\newcommand{\mA}{\mathcal{A}}

\newcommand{\bbP}{\mathbb{P}}

\newcommand{\id}{\mathrm{id}}
\newcommand{\C}{\mathbb{C}}
\newcommand{\F}{\mathbb{F}}

\newcommand{\eop}{\hspace*{\fill}$\blacksquare$}

\newcommand{\btt}{\begin{ttheorem}}
\newcommand{\ett}{\end{ttheorem}}

\newcommand{\bt}{\begin{theorem}}
\newcommand{\et}{\end{theorem}}
\newcommand{\bcc}{\begin{conjecture}}
\newcommand{\ecc}{\end{conjecture}}
\newcommand{\bc}{\begin{corollary}}
\newcommand{\bl}{\begin{lemma}}
\newcommand{\ec}{\end{corollary}}
\newcommand{\el}{\end{lemma}}
\newcommand{\bq}{\begin{question}}
\newcommand{\eq}{\end{question}}
\newcommand{\br}{\begin{remark}}
\newcommand{\er}{\end{remark}}
\newcommand{\bd}{\begin{definition}}
\newcommand{\ed}{\end{definition}}

\newcommand{\U}{\ensuremath{\mathbf{U}}}

\newcommand{\GL}{\ensuremath{\mathbf{GL}}}

\newcommand{\Aut}{\ensuremath{\texttt{Aut}}}

\newcommand{\mH}{\ensuremath{\mathcal{H}}}

\newcommand{\wS}{\mathbf{S}}

\newcommand{\hU}{\mathbf{U}}
\newcommand{\R}{\mathbb{R}}

\newcommand{\supp}{\texttt{supp}}
\newcommand{\Fr}{\texttt{Fr}}

\author{Koen  Thas}

\address{Department of Mathematics,
Ghent University,
Krijgslaan 281, S25, B-9000 Ghent, Belgium}

\email{koen.thas@gmail.com}

\title[Absolute Quantum Theory]{Absolute Quantum Theory (after Chang, Lewis, Minic and Takeuchi),\\ and a road to quantum deletion}

\date{July 2018}

\begin{document}

\maketitle

\begin{abstract}
In a recent paper \cite{chang}, Chang et al. have proposed studying ``Quantum $\F_{un}$'': the $q \mapsto 1$ limit of Modal Quantum Theories over finite fields $\F_q$, motivated by the fact that 
such limit theories can be naturally interpreted in classical Quantum Theory. In this letter, we first make a number of rectifications of statements made in \cite{chang}. 
For instance, we show that Quantum Theory over $\F_1$ {\em does} have a natural analogon of an inner product, and so orthogonality is a well-defined notion, contrary 
to what is claimed in \cite{chang}. Starting from that formalism, we introduce time evolution operators and observables in Quantum $\F_{un}$, and we determine the 
corresponding unitary group. Next, we obtain a typical no-cloning in the general realm of Quantum $\F_{un}$. Finally, we obtain a no-deletion result as well. Remarkably, we show 
that we {\em can} perform quantum deletion by {\em almost unitary operators}, with a probability tending to $1$. Although we develop the construction in Quantum $\F_{un}$, 
it is also valid in any other Quantum Theory (and thus also in classical Quantum Theory).  
\end{abstract}

\medskip
\section{Quantum $\F_{un}$, and $\F_{1^\ell}$}

\subsection{}
In many papers, alternative Quantum Theories have been proposed for classical Quantum Theory (in complex Hilbert spaces, following the K\o benhavn interpretation). For 
instance, there are a number of papers on ``Modal Quantum Theories'' (MQTs), which consider similar theories over finite fields (see e. g. \cite{MQT,Lev}). Whether the motivation is that these 
simply serve as toy models for the classical theory, or that they maybe come closer to physical reality, is arguable. But that fundamental results such as no-cloning 
can also be obtained in MQTs, makes the latter interesting in their own right. \\

\subsection{}
In the last ten years, there has been an increasing interest in the {\em field with one element}; this nonexisting object is contained in every field, and its geometric theory (in a broad sense of  
the word: Algebraic Geometry, Incidence Geometry, ...) is an ``absolute theory'' which is present in any geometric theory over a field. We refer to the monograph \cite{AAFun} for a thorough introduction. A very simple and equally important manifestation of ``$\F_1$'' is the following. Consider the class of all combinatorial projective spaces $\bbP^n(\F_q)$ over finite fields $\F_q$ \cite{Hirsch,Ernie}; 
each such space has an automorphism group $\mathbf{P\Gamma L}_{n + 1}(\F_q)$. Each such space has (A) $q + 1$ points per line, (B) any two different points are contained in precisely one line, and (C) any two different intersecting lines are contained in one axiomatic projective plane of order $q$. Axiomatic projective planes are characterized by three simple properties: (1) property (B); 
(2) any two different lines intersect in precisely one point and (3) there exist four points with no three of them on the same line. Each such plane has an {\em order}: a positive integer $c$ such that each line contains $c + 1$ points and each point is on $c + 1$ lines. If we imagine that $c$ \ul{goes to one}, we end up with a set of points in which each line has two points, and for which (1) and (2) hold. It is easy to see that the set must have three points, and that we obtain the geometry of a triangle. So (3) does not hold anymore. Turning back to the combinatorial geometry of $\bbP^n(\F_q)$ and letting \ul{$q$ go to $1$} (so that we shrink $\F_q$ to a ``field with one element,'' $\F_1$), we end up with a limit geometry in which
\begin{itemize}
\item[(A$'$)]
each line has $2$ points;
\item[(B$'$)]
any two different points are contained in one unique line;
\item[(C$'$)]
each two different intersecting lines are contained in a unique triangle.
\end{itemize}

Obviously, (C$'$) follows from (A$'$) and (B$'$). And it also clear that $\bbP^n(\F_1)$ is a complete graph. {\em Exercise}: observe that its number of points is $n + 1$. The picture only becomes complete after the observation that indeed, a complete graph on $n + 1$ points is a subgeometry of {\em any} combinatorial projective space $\bbP^n(k)$, where $k$ is a field (or even a {\em division ring}, cf. section \ref{GGQT}), and that the group which is induced by $\mathbf{P\Gamma L}_{n + 1}(k)$ on such a subgeometry, is isomorphic to the full symmetric group on $n + 1$ letters. \\

\subsection{}
In \cite{chang}, the authors propose to apply the same formalism on the level of Quantum Theory, so as to interpret phenomena in Modal Quantum Theories in classical Quantum Theory over the complex numbers. 
So \cite{chang} bids for a transition from the finite MQTs to classical Quantum Theory (which we abbreviate by ``AQT,'' referring to ``Actual Quantum Theory'' as in \cite{MQT}) through the limit $q \mapsto 1$. 

\begin{equation}
\label{eqQFun}
\mathbf{MQT}_q\ \overset{q \mapsto 1}{\longrightarrow}\ \mathbf{AQT}.
\end{equation}

This idea is the starting point of the present note. \\

\subsection{}

In \cite{GQT}, we have introduced a general approach to Quantum Theories | in the K\o benhavn setting | over so-called {\em division rings} (we will recall the basics in the next section); 
this approach unifies all known Quantum Theories in this setting, but it also argues that even over the complex numbers, there are very interesting alternate Quantum Theories to the classical one. 
If we want the combinatorics of projective wave space available, we also argued in \cite{GQT} that the approach of General Quantum  Theories (GQTs) is the most general possible. Many other results are 
obtained: for instance, we have showed that no-cloning holds in every GQT. We also showed how to use the ``quantum kernel,'' a singular object which arises from 
the equation which defines the Hermitian form which replaces the inner product in these theories, in both the new and classical theories (e.g., on the level of quantum codes). \\

A different, more general, formulation of the diagram (\ref{eqQFun}) could also be: ``Can one describe a Quantum Theory which ``sees''  (fundamental aspects in) all (Actual, Modal, General) Quantum Theories?'' 
In \cite{GQT} we have introduced such an ``absolute Quantum Theory'' in characteristic $0$: the {\em minimal standard model}, which is defined over the rationals $\mathbb{Q}$. The 
philosophy of minimal models fits very well in the contents of \cite{chang}, and the present paper.

\subsection{A virtual deletion machine in all Quantum Theories}
\label{virtual}

The principle of {\em superposition} is a fundamental property in Quantum Mechanics; if two evolving states $\vert s \rangle_1$ and $\vert s \rangle_2$  solve the Schr\"{o}dinger equation, then an arbitrary linear  combination $a\vert s\rangle_1 + b\vert s\rangle_2$ also is a solution. The famous no-cloning result of Wootters and Zurek \cite{WZclone} and Dieks \cite{Dieks} has been obtained as an implication of the superposition principle, and so has the no-deletion principle of Pati and Braunstein \cite{nodelete}. In General Quantum Theories, the author has shown that 
both no-theorems still hold, and superposition remains to be a key in the proofs \cite{GQT}. 

As $\F_1$-theory lacks addition on the algebraic level (see section \ref{form1}), a major basic question is whether similar no-cloning and no-deletion results will still hold 
in Quantum $\F_{un}$. And whether the diagram (\ref{eqQFun}) remains to have a meaning in the context of such more advanced questions. In the theory of Chang et al. \cite{chang}, such 
questions make no sense, since they have no unitary operators available, but we do. And as we will see, the lack of flexibility due to not having addition at hand, will be compensated 
by the fact that the unitary groups in Quantum $\F_{un}$ are of a restricted type. At the end, we will obtain the no-cloning and no-deletion theorems in Quantum $\F_{un}$. 
 
On the other hand, after introducing {\em almost unitary operators} (which are 
allowed to be singular), we obtain a quantum deletion theory {\em which deletes one copy of any two given state rays with a probability tending to $1$}. The diagram (\ref{eqQFun}) {\em does apply} to this result, so that we virtually obtain {\em deletion} in classical Quantum Theory!

\subsection{Overview}

In this letter, we first make a number of rectifications of statements made in the interesting recent note \cite{chang}. 
For instance, we show that Quantum Theory over $\F_1$ {\em does} have a natural analogon of an inner product, and so orthogonality is a well-defined notion, contrary 
to what is claimed in \cite{chang}. A general and widespread misconception in Modal Quantum theory papers is the common belief that such theories do not allow inproducts (see for instance \cite{MQT}). As explained in \cite{GQT} | see also the next section | this is not true: even in the setting of  general Quantum Theories,  one has natural generalized versions of inproducts available, and in many cases (such as in the case of general Quantum Theories over algebraically closed fields in characteristic $0$), the theory comes with a Born rule as well. 
Starting from that new formalism, we introduce time evolution operators and observables in Quantum $\F_{un}$, and we determine the 
corresponding unitary group. Finally, we develop a no-cloning and no-deleting theory in Quantum $\F_{un}$.

In the next section we tersely review the viewpoint of General Quantum Theory. The following two sections \ref{form1} and \ref{big} prepare in some detail the theory of Quantum $\F_{un}$. 
This includes the completion of what is described in \cite{chang}, but also other aspects which are needed to understand the setting. Section \ref{dic} contains a small 
dictionary which compares some basic aspects of Actual, Modal, General and Absolute Quantum Theories.

With that dictionary in mind, Quantum Information theorists might want to skip section \ref{big}, and focus on sections \ref{noclone} and \ref{nodel}, which form the core of this paper on the level of 
physical applications.

\medskip
\section{A quick review of General Quantum Theory}
\label{GGQT}

In this section, a {\em division ring} is a field in which multiplication not necessarily is commutative.  Example: the quaternions. If one constructs 
a projective space from a left or right vector space over a division ring in the usual way, then one obtains a space of which the underlying combinatorial incidence geometry (which 
one defines by taking the points, lines, planes, etc. of the space, endowed with the natural symmetrized containment relation) still is an {\em axiomatic 
projective space} (in the sense of Veblen and Young \cite{VY}), and division rings are the most general algebraic objects with this property \cite{VY}: the paper \cite{VY} shows that 
axiomatic combinatorial projective spaces of dimension at least three, are projective spaces coming from vector spaces over division rings.

\medskip
\subsection{$(\sigma,1)$-Hermitian forms}
\label{Herm}

Let $k$ be a division ring. An {\em anti-automorphism} of $k$ is a map $\gamma: k \mapsto k$ such that
$\gamma$ is bijective;
for any $u, v \in k$, we have $\gamma(u + v) = \gamma(u) + \gamma(v)$;
and for any $a, b \in k$, we have $\gamma(ab) = \gamma(b)\gamma(a)$.

If $k$ is a commutative field, then anti-automorphisms and automorphisms coincide. 
Note that the fields $\mathbb{Q}$ and $\mathbb{R}$ do not admit nontrivial automorphisms.

\subsection{Hermitian forms}

Suppose that $k$ is a division ring, and suppose $\sigma$ is an anti-automorphism of $k$. Let $V$ be a right vector space over $k$.
A {\em $\sigma$-sesquilinear form} on $V$ is a map $\nu: V \times V \mapsto k$ for which we have the following properties: 
\begin{itemize}
\item
for all $a,b,c,d \in V$ we have that $\nu(a + b,c + d) = \nu(a,c) + \nu(b,c) + \nu(a,d) + \nu(b,d)$;
\item
for all $a, b \in V$ and $\alpha, \beta \in k$, we have that $\nu(a\alpha,b\beta) = \sigma(\alpha)\nu(a,b)\beta$.
\end{itemize}

We have that $\nu$ is reflexive if and only if  there exists an $\epsilon \in k$ such that for all $a, b \in V$, we have
\begin{equation}
\nu(b,a) = \sigma\Big(\nu(a,b)\Big)\epsilon.
\end{equation}

Such sesquilinear forms are called {\em $(\sigma,\epsilon)$-Hermitian}. If $\epsilon = 1$ and $\sigma^2 = \id \ne \sigma$, then we speak of 
a {\em Hermitian form}. 

The standard inner product (and in fact {\em any} inner product) in a classical Hilbert space over $\C$ is a Hermitian form.

\medskip
\subsection{Standard $(\sigma,1)$-Hermitian forms}

If $k$ is a division ring with involution $\sigma$, the {\em standard $(\sigma,1)$-Hermitian form} on the right vector space $V(d,k)$, is given by

\begin{equation}
\Big\langle x \Big\vert y \Big\rangle\ :=\ x_1^\sigma y_1 + \cdots + x_d^\sigma y_d, 
\end{equation}
where $x = (x_1,\ldots,x_d)$ and $y = (y_1,\ldots,y_d)$. 

In the case that $\sigma = \id$, we obtain a form which is usually called {\em symmetric}; it is not a proper Hermitian form, but still comes in handy in some situations
(for example in cases of field reduction: ``real Hilbert spaces'' have often been considered in Quantum Theory; see e.g. \cite{Wootreal,Wootreal2,Wootreal3}).

\subsection{The unitary group $\U(V,\varphi)$}
\label{morph}

An {\em automorphism} of a $(\sigma,1)$-Hermitian form $\varphi$ on the $k$-vector space $V$, is a bijective linear operator $\omega: V \mapsto V$
which preserves $\varphi$, that is, for which 
\begin{equation}
\varphi(\omega(x),\omega(y)) = \varphi(x,y)
\end{equation}
for all $(x,y) \in V \times V$. The group of all such automorphisms is called the {\em unitary group}, and denoted $\U(V,\varphi)$. 

Example. Let $k = \C$, $\sigma$ be complex conjugation, and $V = V(n,\C)$. Then $\hU(V,\varphi) = \mathbf{GU}_n(\C) = \mathbf{U}(n)$.

\medskip
\subsection{GQT}

If we speak of ``division ring with involution,'' we mean a division ring with an involutory anti-automorphism.\\

$\left\vert
\begin{tabular}{p{.9\textwidth}}
From now on, we propose to depict a physical quantum system by a \ul{general Hilbert space $\mH = \Big((V(\omega,k),+,\cdot),\langle \cdot,\cdot \rangle\Big)$, with $k$ a division ring with involution $\sigma$, and $\Big\langle \cdot \Big\vert\cdot \Big\rangle$ a $(\sigma,1)$-Hermitian form.} \\
\end{tabular}
\right.$

\bigskip
If we speak of ``standard GQT,'' we mean that given $\sigma$, the general Hilbert space comes with the standard $(\sigma,1)$-Hermitian form. Also, as some fields such as 
the reals and the rational numbers do not admit nontrivial involutions, they only can describe ``improper'' quantum systems. By extension of Quantum Theories (which is described in \cite{GQT}), this is no problem (as often has been the case when switching between AQT over $\C$ and $\R$).

In this paper, we will only focus on the standard Hermitian forms, to keep the analogy with AQT as clear as possible. We also refer to section \ref{dic} for an overview of some 
basic notions in the different Quantum Theories.

\medskip
\section{The formalism over $\F_1$}
\label{form1}

\medskip
\subsection{$\F_1$ and $\F_{1^2}$}

As in \cite{chang}, we define $\F_1$ to be the set $\{ 0,1\}$ endowed with the obvious multiplication, once we agree that $0$ is the absorbing element. 
So $0 \cdot 1 = 1 \cdot 0 = 0 \cdot 0 = 0$, and $1 \cdot 1 = 1$. We adopt the traditional \ul{$\F_1$-Mantra} that {\em there is no addition} \cite{Funext}. 

Define the quadratic extension of $\F_1$, and denoted by $\F_{1^2}$, as the set $\{0\} \cup \mu_2$, where $\mu_2$ is the group of two elements, 
multiplicatively written as $\{ 1,a\}$. So $a \cdot 1 = 1 \cdot a = a$ and $a^2 = 1$. Again, $0$ is the absorbing element. We define 
$\F_{1^\ell}$, for any positive integer $\ell$, in a similar manner (replacing $\mu_2$ by the cyclic group $\mu_\ell$ of order $\ell$).

\medskip
\subsection{Vector space over $\F_{1^\ell}$, and the affine viewpoint: frames}
\label{frame}

Fix a dimension $m$ (positive integer different from $0$). In this section, our state space will be the vector space $V= V(m,\F_{1^\ell})$. We will take on the viewpoint 
of our recent paper \cite{Funext}, and also, we will make no distinction at this point between vector spaces over $\F_1$ and its extensions, and affine spaces. 

Let $S$ be $\F_1$, or $\F_{1^2}$. The {\em affine frame} $\mA(m,S)$ is defined as  the set 
\begin{equation}
\{ (s_1,\ldots,s_m)\ \vert\ s_i \in S \}. 
\end{equation}

Let $(0,\ldots,0) =: \omega$. In the case $S = \F_{1^\ell}$, we will use the same definitions. 

Besides $\omega$, the frame points with exactly {\em one nonzero entry} play a special role; those were the points used in \cite{chang} in the case 
$S = \F_1$. We call such points {\em simple points}. 
As we showed in \cite{Funext}, and as was conjectured by others (see the details in \cite{Funext}), we need to include the extra points 
to set up a natural connection with the ``functor-of-points viewpoint.''  

It is very important to note that \ul{we cannot add the points in a frame}, as the underlying set $S$ only is endowed with multiplication. \\

In the rest of this section, we solely work over $\F_{1^2}$ to fix ideas. 

\medskip
\subsection{The standard form}
\label{stan}

As we will later see, in section \ref{big}, $\F_{1^2}$ does not allow a nontrivial involution. On the other hand, there is a natural standard $(1,1)$-Hermitian 
form, being:
\begin{equation}
\label{stand}
\Big\langle \overline{x} \Big\vert \overline{y} \Big\rangle  := x_1y_1 + \cdots + x_my_m = \overline{x}^T\cdot\overline{y}. 
\end{equation}

Now the $\F_1$-Mantra adds an extra rule to the formalism: in (\ref{stand}), $\Big\langle \overline{x} \Big\vert \overline{y} \Big\rangle$ only has a meaning if 
{\em at most one nonzero} occurs in the summation in (\ref{stand}). 

In the next subsection, we will look at the implications for orthogonality relations.

\medskip
\subsection{Orthogonality}
\label{orth}

If $\overline{x} = (x_1,\ldots,x_m)$ is a point in a vector space over $\F_{1^\ell}$, then by $\supp(\overline{x})$ we denote the set of indices $j$ for which $x_j \ne 0$ (``support'' of  $\overline{x}$). 
By $\supp^c(\overline{x})$, we denote the complement $\{ 1,\ldots,m \} \setminus \supp(\overline{x})$. Then vectors $\overline{u}$ and $\overline{v}$ are orthogonal if and only if 
\begin{equation}
\supp(\overline{u}) \subseteq \supp^c(\overline{v}).
\end{equation}

If we now define $\overline{x}^\perp$ as 
\begin{equation}
\Big\{ \overline{y}\ \Big\vert\ \Big\langle \overline{x} \Big\vert \overline{y} \Big\rangle = 0 \Big\},
\end{equation}
then the dimension of these spaces obviously (only) depends on $\vert \supp(\overline{x}) \vert$ | in vector spaces over a field such as $\C$, this dependence is not present. 
Over $\F_{1^\ell}$ we have that $\overline{x}^\perp$ is a vector space of dimension $\vert \supp^c(\overline{x}) \vert$; for example, if $\overline{x}$ is a simple point, then $\overline{x}^\perp$ is a hyperplane (dimension $m - 1$, so classical behavior), and if $\overline{x}$ has maximal support $m$, then $\overline{x}^\perp$ is $0$-dimensional, only consisting of the zero vector.

\medskip
\subsection{Time evolution and Hermitian operators}
\label{time}

Before defining time evolution and Hermitian operators, we need to know what {\em linear operators} of $V(m,\F_{1^2})$ are. 
As $\F_1$-theory, and in particular the vector spaces/frames as defined in subsection \ref{frame},  does not allow addition, we can only work with matrices with entries in $\F_{1^2}$ which have at most 
one nonzero entry in each row and column. If we want these operators to be invertible, we ask that every row and column 
precisely has {\em one} nonzero entry. This means that every such $(m \times m)$-matrix has the structure of a permutation matrix, but the nonzero entries vary over 
$\mu_2$. 
As the set of all $(m \times m)$-permutation matrices forms a group which is isomorphic to the symmetric group $\wS_m$ on $m$ letters, we can thus write that 
\begin{equation}
\GL(m,\F_{1^2}) := \mu_2 \wr \wS_m,\ \ \mbox{a generalized symmetric group $S(2,m)$};
\end{equation}
such groups are also known as ``signed symmetric groups.''

An element $A \in \GL(m,\F_{1^2})$  preserves the standard form if and only if $A^TA = \id_m$.  Every permutation matrix satisfies this identity, and since the entries of elements of 
$\GL(m,\F_{1^2})$ are contained in $\F_{1^2}$, this property remains to be true; so we obtain that 
\begin{equation}
\mathbf{U}(m,\F_{1^2}) = \GL(m,\F_{1^2}).
\end{equation}

This is not true for general extensions $\F_{1^\ell}$, but by accident, \ul{the unitary groups over $\F_{1^2}$ are maximally large}. 

If we now adapt the notion of observable, we obtain that $H \in \GL(m,\F_{1^2})$ defines an observable if and only if $H^T = H$, that is, if and only if 
\begin{equation}
H^2 = \id_m. 
\end{equation}

Again, the outcome is incidental because we are working over $\F_{1^2}$!

\medskip
\section{The bigger picture}
\label{big}

Now that we have introduced the basics of a K\o benhavn Quantum Theory over $\F_1$, it is necessary to \ul{extend the theory to 
arbitrary extensions of $\F_1$}. This consideration yields extra elbow room for theory and applications, as we will see.

\subsection{The Frobenius maps}
\label{Frob}

Let $\overline{\F_1}$ be the {\em algebraic closure} of $\F_1$; it consists of all complex roots of unity plus an element $0$, endowed with the natural 
multiplication; see \cite{Funext,motive}. For every positive integer $\ell$, we have that $\F_{1^\ell} \leq \overline{\F_1}$. Elements of $\F_{1^\ell}$ are characterized by the fact that they 
are precisely the solutions of the equation
\begin{equation}
x^{\ell + 1} = x.
\end{equation}

Compare this to the analogous situation for the algebraic closure $\overline{\F_q}$ of the finite field $\F_q$; in this case, the Frobenius map $\Fr^q: v \mapsto v^q$ singles out the elements 
of $\F_q$. 

Following \cite{Funext,motive}, we call the map $\Fr_1^{\ell + 1}: \overline{\F_1} \mapsto \overline{\F_1}: u \mapsto u^{\ell + 1}$ the {\em absolute (or $\F_1$-) Frobenius endomorphism} of 
degree $\ell + 1$. We use the same name if the domain of $\Fr_1^{\ell + 1}$ is reduced. \\

\subsection{$\Aut(\F_{1^\ell})$}

An {\em automorphism} of $\F_{1^\ell}$ is a permutation $\varphi$ of $\F_{1^\ell}$ such that $\varphi(ab) = \varphi(a)\varphi(b)$ for all $a, b$. Note that $\varphi(0) = 0$ and $\varphi(1) = 1$. 
The set of all automorphisms 
of $\F_{1^\ell}$ is denoted by $\Aut(\F_{1^\ell})$, and is a group if we endow it with the group law ``composition of maps.''

Note that all automorphisms of $\F_{1^\ell}$ are given by maps $u \mapsto u^m$, with $(\ell,m) = 1$ (as they correspond to automorphisms of the cyclic group $\mu_\ell$). It follows that 
$\Aut(\F_{1^\ell})$ is isomorphic to the group of multiplicative units in the ring $\mathbb{Z}/\ell\mathbb{Z}$.

\subsection{Involutions of the fields $\F_{1^\ell}$}

The involutory automorphism ``complex conjugation'' plays a crucial role in Actual Quantum Theory over $\C$ (for instance, to define the standard inner product, orthogonality, etc.), and 
in GQT, an analogous role is played by the involutory (anti-)automorphisms.  
We need to understand such maps in the context of Quantum Theories over $\F_{1^\ell}$.

The following lemma classifies involutions of extension of $\F_1$. 

\begin{lemma}
\label{classinv}
The map $\mathrm{\Fr}_1^{r + 1}$ defines a nontrivial involutory automorphism of $\F_{1^m}$ if and only if the following conditions are satisfied:
\begin{itemize}
\item[SUB]
$m$ divides $r(r + 2)$;
\item[NTRIV]
$m$ does not divide $r$. 
\end{itemize}
\end{lemma}

{\em Proof}.\quad
Let $\sigma: v \mapsto v^{r + 1}$ be a nontrivial involutory automorphism of $\F_{1^m}$. Then 
\begin{equation}
\label{eqinv}
u^{(r + 1)(r + 1)} = u 
\end{equation}
for all $u \in \F_{1^m}$. If we pass to $\overline{\F_1}$, then 
all solutions of (\ref{eqinv}) are precisely given by the elements of $\F_{1^{r(r + 2)}}$ (see subsection \ref{Frob}), so (SUB) holds. On the other hand, the fixed field of $\sigma$ 
in $\F_{1^{r(r + 2)}}$ is $\F_{1^r}$, so since we assume $\sigma$ to be nontrivial, $m$ cannot divide $r$ (NTRIV): $m$ divides $r$ if and only if $\mu_m$ is a subgroup of 
$\mu_r$ if and only if $\F_{1^m}$ is a subfield of $\F_{1^r}$.

Finally, for $\sigma$ to be an automorphism of $\F_{1^m}$, we need to invoke the necessary and sufficient condition (AUT):  $(r + 1,m) = 1$; this property follows from (SUB).
\eop \\

{\em Examples}. The map $\Fr_1^{r + 1}: v \mapsto v^{r + 1}$ induces involutions in the following natural cases: $\F_{1^{r + 2}}$, $\F_{1^{2r}}$ (in case $r$ is even), $\F_{1^{2(r + 2)}}$ (in case $r$ is even), and $\F_{1^{r(r + 2)}}$. The latter example might seem more natural to some, if one replaces $r$ by $\ell - 1$; we then get that the absolute Frobenius $\Fr_1^\ell$ is an involutory 
automorphism of $\F_{1^{\ell^2 - 1}}$ (which has size $\ell^2)$, with fixed field $\F_{1^{\ell - 1}}$ (which has size $\ell$). This strongly resembles the case of finite fields when $\ell$ is a prime power.

\subsection{Standard examples of Absolute Quantum Theories}

In this subsection, we describe an additional Absolute Quantum Theory with standard inproduct, based on our knowledge of Lemma \ref{classinv} (and the examples following the lemma). In a similar way, one can describe all Absolute Quantum Theories.

So we consider the field $\F_{1^{r(r + 2)}}$ and its absolute Frobenius automorphism $\Fr_1^{r + 1}: v \mapsto v^{r + 1}$. By Lemma \ref{classinv} we know that it is 
involutory, and the fixed field is $\F_{1^r}$.  We represent our states in the state space $V(n,\F_{1^{r(r + 2)}})$, with standard inproduct

\begin{equation}
\label{stand2}
\Big\langle \overline{x} \Big\vert \overline{y} \Big\rangle = x_1^{r + 1}y_1 + \cdots + x_n^{r + 1}y_n. 
\end{equation}

%The quantum kernel is defined by the 

Observables are Hermitian operators $H$ satisfying 
\begin{equation}
H = \Big{[H^T\Big]}^{\Fr_1^{r + 1}}; 
\end{equation}
the underlying structure is that of an involutory permutation matrix, and for each nonzero entry 
$h_{ij}$, we must have that 
\begin{equation}
h_{ij} = h_{ji}^{r + 1}. 
\end{equation}
Note that $h_{ij} = h_{ji}^{r + 1}$ if and only if $h_{ji} = h_{ij}^{r + 1}$ since $\Fr_1^{r + 1}$ is an involution. 

Also note that all symmetric matrices in $\GL(n,\F_{1^{r(r + 2)}})$ with only entries in $\F_{1^r}$ satisfy this condition.   \\

Unitary operators are given by operators $U$ for which 
\begin{equation}
\Big{[U^T\Big]}^{\Fr_1^{r + 1}}U = \id_m; 
\end{equation}
every permutation matrix satisfies this identity, and for $U$ defined over 
$\F_{1^{r(r + 2)}}$, we must have that each nonzero entry $a$ satisfies  $a^{r + 2} = 1$, that is, $a \in \F_{1^{r + 2}}$. 

\bt[Unitaries and observables]
With $\sigma = \Fr_1^{r + 1}$, $\mathbf{U}(m,\F_{1^{r(r + 2)}})$ is given by the wreath product $\mu_{1^{r + 2}} \wr \mathbf{S}_m$, the generalized symmetric group $S(r + 2,m)$. And observables are given by $(m \times m)$-matrices 
with precisely one nonzero element in each row and column, for which $h_{ij} = h_{ji}^{r + 1}$ for each nonzero entry $h_{ij}$. \eop
\et

The unitaries and observables now look very different than those in subsection \ref{time}!

\medskip
\subsection{Orthogonality}

If we work in Quantum Theories over extensions of type $\F_{1^\ell}$, everything we have observed in subsection \ref{orth} remains valid.

\medskip
\section{Dictionary}
\label{dic}

In the instructive table below, we compare Actual Quantum Theory, Modal Quantum Theory, General Quantum Theories and Absolute Quantum Theory. For the latter, we only have plugged in the last example 
of the previous section. In the case of Modal Quantum Theory, we present its ``completed version'' which I described in \cite{GQT}.
Also, $k$ denotes the algebraic structure over which we coordinatize Hilbert spaces; $\sigma$ is an involutory automorphism of $k$; 
$k_{\sigma} := \{ a \ \vert\ a^\sigma = a \}$ plus induced field structure of $k$; and the last column provides the standard ``inner product'' defined by $\sigma$.  Note that $k_\sigma$ 
plays an important role in the formulation of \ul{Born's rule}; for instance, in Actual Quantum Theory over $\C$, we consider a quantum system described by the wave 
function $\vert \xi \rangle$, and suppose $\vert \phi_i \rangle$ is an eigenvector of an orthogonal base of eigenvectors of an observable $H$; also, let $\lambda_i$ be the 
corresponding eigenvalue. If we write $\langle \phi_i \vert \xi \rangle = c + id$, with $c, d \in \mathbb{R}$, then the probability of the measurement $\lambda_i$ is 
\begin{equation}
\Big\vert \langle \phi_i \vert \xi \rangle \Big\vert^2 = c^2 + d^2.  
\end{equation}

As we mentioned in \cite{GQT}, one can formulate this rule in many other GQTs.

\begin{table}[h]
\begin{tabular}{|c|c|c|c|c|}\hline
Quantum Theory & $k$ & $\sigma$ & $k_{\sigma}$  &standard form $\Big\langle \overline{x} \Big\vert \overline{y} \Big\rangle $\\ \hline
&&&&\\
 Actual Quantum Theory & $\C$ & $v \mapsto \overline{v}$ &$\mathbb{R}$   &$\overline{x_1}y_1 + \cdots + \overline{x_m}y_m$    \\
 &&&& \\ \hline
 &&&&\\
 Modal Quantum Theory & $\F_{q^2}$ & $v \mapsto v^q$ &$\F_q$   &$x_1^qy_1 + \cdots + x_m^qy_m$    \\
 &&&& \\ \hline
 &&&&\\
 General Quantum Theory & division ring with involution $\sigma$ & $\sigma$ &$k_\sigma$   &$x_1^{\sigma}y_1 + \cdots + x_m^{\sigma}y_m$     \\
&&&&\\  \hline
 &&&&\\
 Absolute Quantum Theory & $\F_{1^{r(r + 2)}}$ & $v \mapsto v^{r + 1}$ &$\F_{1^r}$   &$x_1^{r + 1}y_1 + \cdots + x_m^{r + 1}y_m$     \\
&&&&\\  \hline
%  one & two & three \\\hline
%  one & two & three \\\hline
\end{tabular}
\end{table}

\medskip
\section{One cannot clone an unknown state in Absolute Quantum Theory}
\label{noclone}

A result of Wootters and Zurek \cite{WZclone,WZclone2}, and independently Dieks \cite{Dieks}, states that one cannot ``clone'' an unknown state. Formally, one wants to solve the next equation:
\begin{equation}
\label{cloneq}
U\cdot \Big(\vert \phi \rangle_A \otimes \vert e \rangle_B\Big) = \vert \phi \rangle_A \otimes  \vert \phi \rangle_B
\end{equation}
where $\vert \phi \rangle_A$ is an unknown state in a complex Hilbert space $H_A$ and $\vert \phi \rangle_B$ is the clone in the Hilbert space $H_B$ (which is 
a copy of $H_A$), where $\vert e \rangle_B$ is an unknown blank state in $H_B$, and $U$ is a unitary operator. 

Since $\vert \phi \rangle_A$ is arbitrary, we can replace it by a linear combination 
\begin{equation}
\label{sum}
\alpha\vert \phi \rangle_A + \beta \vert \phi' \rangle_A, 
\end{equation}
and 
then the unitarity of $U$ (or better: the {\em linearity}) easily leads to a contradiction. We refer to the discussion in subsection \ref{virtual} for additional remarks.

In \cite{MQT}, the authors have shown that similarly, one cannot clone an unknown state in Modal Quantum Theory over prime fields. 
In \cite{GQT}, we obtained the general result that one cannot clone an unknown state in a General Quantum Theory over {\em any} division ring. Due to the degree 
of generality, the proof of that result is slightly more subtle than the complex or modal case.\\

Over $\F_{1^\ell}$, one cannot use mixed states such as (\ref{sum}), since we cannot {\em add} states. So we need a (slightly) different approach. 
We will consider (\ref{cloneq}) with only pure states, and take $U$ to be in $\mathbf{U}(m,\F_{1^\ell})$. We will also suppose that $\vert e \rangle_B$ 
is an unknown, but fixed, blank state. As we will see, the particular nature of unitary operators over $\F_{1^\ell}$ already prevents the fact that the 
blank state can be randomly chosen. 

First of all, note that the following identity should hold for any state $\vert \phi \rangle_A$ and any $\alpha \in \F_{1^2}$:
\begin{equation}
\label{prod}
U\Big(\alpha\vert \phi \rangle_A \otimes \vert e \rangle_B\Big) = \alpha \Big(\vert \phi \rangle_A \otimes  \vert \phi \rangle_B\Big) = \alpha\vert \phi \rangle_A \otimes  \alpha\vert \phi \rangle_B,
\end{equation}
so that $\alpha^2 = \alpha$ for all $\alpha$, which is already false, even for simple states. \\

In what follows, we will therefore work on the {\em projective} level, to see what the influence of factors is in this context. 

We first work with an unknown {\em simple} state $\vert \phi \rangle_A$. As $U$ must have the structure of a permutation matrix, the fact that 
 $\vert \phi \rangle_A \otimes \vert \phi \rangle_B$ is simple, implies that $\vert e \rangle_B$ also must be simple. 
 But then if we consider a state $\vert \phi' \rangle_A$ which is {\em not} simple, obviously the identity (\ref{cloneq}) cannot work, due again to 
 the permutation matrix structure of $U$. 
 
 Still, as states are only determined up to factors, the 
 natural question arises whether we can clone, \ul{projectively, the simple state rays}. This is in fact very easy: as we have seen, since we are only cloning 
 simple states, $\vert e \rangle_B$ must be a simple state itself. On the other hand, since we are working in a projective space, there are only 
 $m$ different simple states if we assume that $H_A$ has dimension $m$ over $\F_{1^\ell}$; in fact, if $H_A$ would have dimension $m$ over $\F_1$, we would obtain 
 essentially the same points. Obviously, we can find a permutation matrix $U$ in $\mathbf{U}(m,\F_{1^\ell})$ which maps 
$\vert \phi \rangle_A \otimes \vert e \rangle_B$ to $\vert \phi \rangle_A \otimes  \vert \phi \rangle_B$ with $\vert \phi \rangle_A$ varying through the set of 
simple states, so that we obtain a ``simple cloning'' result. 

Interpreting this result on the level of the classical case (so over $\C$), we obtain the well-known understanding that orthogonal states indeed indeed can be cloned (see 
for instance Wootters and Zurek \cite{WZclone,WZclone2}) (note that in the initial vectorial case, the simple cloning result also work if one assumes the simple states one is considering to be orthogonal). So in the philosophy of Chang et al. \cite{chang}, we obtain a new instance of the formalism

\begin{equation}
\mathbf{MQT}_q\ \overset{q \mapsto 1}{\longrightarrow}\ \mathbf{AQT}.
\end{equation}

\bigskip
\section{Quantum deletion in the absolute and actual context}
\label{nodel}

In \cite{nodelete}, Pati and Braunstein obtain a no-deleting result in Actual Quantum Theory, which was later shown to hold in all GQTs, in \cite{GQT}. 
Formally, one now wants to solve the next equation:
\begin{equation}
\label{deleq}
U\cdot  \Big(\vert \phi \rangle_A \otimes  \vert \phi \rangle_B \Big) =  \vert \phi \rangle_A \otimes \vert e \rangle_B, 
\end{equation}
where $\vert \phi \rangle_A$ is an unknown state in a complex Hilbert space $H_A$ and $\vert \phi \rangle_B$ is the copy of $\vert \phi\rangle_A$ in the Hilbert space $H_B$ (which is 
a copy of $H_A$), where $\vert e \rangle_B$ is an unknown blank state in $H_B$, and $U$ is a unitary operator. \\

The simplest proof of the fact that no such $U$ can exists, seems to be the following: simply observe that if $U$ is as above, then $U^{-1}$ is a cloning operator in the sense of the previous 
section, so that we can finish the proof by using that section. \\

Still, it might be interesting to consider the problem in a more general context, and to allow ``singular unitary operators.'' In fact, {\em because} we can show that quantum deletion is not possible, simply by inverting $U$, this very fact {\em suggests} that the initial definition of quantum deletion of \cite{nodelete} {might not be the correct one}.
Call an operator $U$ (seen as an $(m \times m)$-matrix) {\em almost unitary} if  every nonsingular submatrix which is constructed by deleting columns and rows with the same column --and row indices,  is unitary. Many other alternative definitions could be formulated. In any case, if an almost unitary operator is nonsingular, it is unitary. 

Now we consider the equation (\ref{deleq}) in Absolute Quantum Theory, and allow $U$ to be almost unitary (in the absolute context).  
In exactly the same way as in the previous section, we find that $\alpha^2 = \alpha$ for each $\alpha \in \F_{1^\ell}$. So again, we look at the more natural projective situation. 

Observe that if $\vert \phi \rangle_A$ is a simple state, then $\vert e \rangle_B$ necessarily is simple. We suppose without loss of generality, that 
the first entry of $\vert e \rangle_B$ is $1$, and that the others are $0$. Now define $U$, an $(m^2 \times m^2)$-matrix, as follows:
$\left\vert
\begin{tabular}{p{.8\textwidth}}
\begin{itemize}
\item
$U_{11} = U_{(m + 1)(m + 1)} = \cdots = U_{(m^2 - m + 1)(m^2 - m + 1)} = 1$;
\item
all other entries are $0$.
\end{itemize}
\end{tabular}
\right.$

\medskip
Then obviously, $U$ is almost unitary over $\F_{1^\ell}$, $\C$, and any other division ring/field. 

Now consider any state $\vert \phi \rangle_A = (a_1 \ldots a_m)^T$ with first entry $a_1 \ne 0$. Then, with $\mathbf{0}$ denoting the $\Big((m - 1) \times 1\Big)$-zero matrix, 
\begin{align}
U\cdot  \Big(\vert \phi \rangle_A \otimes  \vert \phi \rangle_B \Big) &= a_1\cdot \begin{pmatrix}      
a_1 \\ \mathbf{0}\\ a_2 \\ \mathbf{0}\\  \vdots \\a_m \\ \mathbf{0}        
\end{pmatrix}
= a_1\cdot \vert \phi\rangle_A \otimes \begin{pmatrix}
1 \\ \mathbf{0}
\end{pmatrix}
= a_1\cdot\vert \phi \rangle_A \otimes \vert e \rangle_B,
\end{align}
and since we work projectively, this means that $U$ \ul{indeed quantum deletes one copy from every such $\vert \phi \rangle_A \otimes  \vert \phi \rangle_B$}. 

If $a_1 = 0$, then $U$ maps $\vert \phi \rangle_A \otimes  \vert \phi \rangle_B$ to the zero element of the vector space | that is, its action on the projective state points with 
$a_1 = 0$ is not defined. As the condition $a_1 \ne 0$ defines an affine subspace of the same dimension as the projective space, and as our considerations do not use 
the fact that we are working over $\F_{1^\ell}$, the formalism is also true for Actual Quantum Theory, and also for every GQT.  
%And of course, we can also switch between  affine space 
%(which is of dimension $m - 1$) and vector space of dimension $m - 1$. 

\begin{theorem}[Quantum deletion by almost unitary operators]
There exists an almost unitary operator $U: \mH_A \otimes \mH_B \mapsto \mH_A \otimes \mH_B$, where $\mH_A$ and $\mH_B$ are copies of the same 
Hilbert space over $\C$, over any division ring with involution, or over $\F_{1^\ell}$, and a blank state $\vert e \rangle_B$, such that $U$ quantum deletes one copy 
in each (projective) state space point $\vert \phi \rangle_A \otimes \vert \phi \rangle_B$ for which the first coordinate is not zero. 
\eop 
\end{theorem}

Consider an $\F_1$-extension $\F_{1^\ell}$, or a finite field $\F_q$ with $\vert \F_q \vert = \ell + 1$. Then the probability that we pick a point 
in the affine subspace $a_1 \ne 0$ in the projective ray state space $\mathbb{P}^{m - 1}(\F_{1^\ell})$ or $\mathbb{P}^{m - 1}(\F_q)$ of $\mH_A$, 
is 
\begin{equation}
\frac{\Big(\ell + 1\Big)^{m - 1}}{\Big((\ell + 1)^{m - 1} - 1 \Big)/\ell} = \frac{\ell}{\ell + (1 - 1/(\ell + 1)^{m - 1})} =: P_{a_1}. 
\end{equation}

We have that 
\begin{equation}
\lim_{m \mapsto \infty}P_{a_1} = \frac{\ell}{\ell + 1},\ \  \mbox{while}\ \ \lim_{\ell \mapsto \infty}P_{a_1} = 1.
 \end{equation}

Now let $\mH_A$ be a vector space over an infinite field or division ring $k$. To fix ideas, one can put $k = \C$. We cannot change a uniform distribution 
on $\mH_A$ (which we identify with the diagonal subspace $\{ \vert \phi \rangle_A \otimes \vert \phi \rangle_B \}$ of $\mH_A \otimes \mH_B$), but on the other hand, it is well known that 
the Lebesgue measure of a hyperplane in a projective space $\mathbb{P}^n(k)$ is zero. We interpret this fact as the idea that the 
probability of choosing a point outside a given hyperplane in $\mathbb{P}^{m - 1}(k)$ tends to $1$.  \\

\medskip
\begin{remark}{\rm
Note that by considering a nonsimple $\vert \phi \rangle_A$ (and a simple $\vert e \rangle_B$ as above), one already sees that cloning is {\em not} possible for 
almost unitary operators as well. 
}
\end{remark}

\end{document}